\documentclass[showpacs,prl,twocolumn,floatfix,
nofootinbib,superscriptaddress]{revtex4}
\usepackage{graphicx}

\newcommand{\be}{\begin{equation}}
\newcommand{\ee}{\end{equation}}

\begin{document}


\title{The $B$ Meson Decay Constant from Unquenched  Lattice QCD}

\author{Alan Gray}
\affiliation{Department of Physics,
The Ohio State University, Columbus, OH 43210, USA}
\author{Matthew Wingate}
\affiliation{Institute for Nuclear Theory, University of Washington,
Seattle, WA 98195-1550, USA}
\author{Christine T.\ H.\ Davies}
\affiliation{Department of Physics \& Astronomy,
University of Glasgow, Glasgow, G12 8QQ, UK}
\author{Emel Gulez}
\affiliation{Department of Physics,
The Ohio State University, Columbus, OH 43210, USA} 
\author{{G.Peter} Lepage}
\affiliation{Laboratory of Elementary Particle Physics,
Cornell University, Ithaca, NY 14853, USA}
\author{Quentin Mason}
\affiliation{Department of Applied Mathematics and Theoretical Physics,
 University of Cambridge, Cambridge, UK}
\author{Matthew Nobes}
\affiliation{Laboratory of Elementary Particle Physics,
Cornell University, Ithaca, NY 14853, USA}
\author{Junko Shigemitsu}
\affiliation{Department of Physics,
The Ohio State University, Columbus, OH 43210, USA}

\collaboration{HPQCD Collaboration}
\noaffiliation


\begin{abstract}
We present  determinations of the $B$ meson decay constant
 $f_B$ and of the ratio $f_{B_s}/f_B$ using the MILC collaboration 
unquenched gauge configurations which include three flavors of light
sea quarks. The mass of one of the sea
quarks is kept around the $strange$ quark mass, and we explore a range 
in masses  for the two lighter sea quarks down to $m_s/8$.
  The heavy $b$ quark is simulated 
using Nonrelativistic QCD, and both the valence and sea
 light quarks are represented by 
the highly improved (AsqTad) staggered quark action.
 The good chiral properties of the latter action allow for a much
smoother chiral extrapolation to physical $up$ and $down$ quarks 
than has been possible in the past.  We find $f_B = 216(9)(19)(4)
(6) \; {\rm MeV}$ and $f_{B_s} /f_B = 1.20(3)(1)$.  
\end{abstract}


\maketitle


Accurate determination of the CKM matrix of the Standard Model and 
tests of its consistency and unitarity constitute an important part of current 
research in experimental and theoretical particle physics.
Experimental studies of neutral $B_d - \overline{B}_d$ mixing,
  carried out as part of this program, are 
now well established and the mass difference $\Delta M_d$ is known 
with high precision \cite{pdg}.  Uncertainty in our present knowledge of 
the CKM matrix element $|V_{td}|$ is hence dominated by theoretical 
uncertainties, the most important of which are errors in $f_B \sqrt{B_B}$, 
where $f_B$ is the $B$ meson decay constant and $B_B$ its bag parameter. 
Lattice QCD allows for first principles calculation of the hadronic matrix 
elements that lead to $f_B$ and $f_B \sqrt{B_B}$ and in recent years the 
onus of reducing theoretical errors in determinations of $|V_{td}|$ has 
been on the Lattice QCD community.  In this article we address and 
significantly improve upon two of the errors that have plagued 
$f_B$ calculations on the lattice in the past, namely uncertainties due to 
lack of correct vacuum polarization in the simulations and 
errors due to chiral extrapolations to physical $up$ and $down$ quarks. 
 The generation of unquenched 
gauge configurations by the MILC collaboration \cite{milc1}, which include 
effects of vacuum polarization from the $strange$ plus 
two lighter dynamical quarks, has led to successful and 
 realistic full QCD calculations 
of a variety of quantities involving both heavy and light quarks 
\cite{milc2,prl,fbsprl,lmass,dsemil,mbc,upsilon}.  Here we also 
take advantage of these well tested configurations.  
Another innovation in recent years has been to use the 
same improved staggered light quark action \cite{stagg}, which is being
 employed for sea light quarks and for light hadron physics, also 
for the valence light quarks inside heavy-light mesons \cite{hlstagg}.
 This has been crucial  for allowing heavy-light simulations close to 
the real world.
Chiral extrapolation requirements are now much milder
than in the past thus reducing effects 
coming from this source of uncertainty.

 In this study we work mainly with four of the ``coarse'' MILC ensembles with 
lattice spacing $a$ around $0.12$fm. 
These have dynamical light quark masses (in units 
of the $strange$ quark mass) of
$m_f/m_s = 0.125, 0.175, 0.25$ and $0.5$.
We have also accumulated results on two of MILC's ``fine'' lattices 
with $a \sim 0.087$fm.
On the fine lattices we use staggered valence light propagators created by
 the Fermilab collaboration. The heavy $b$ quark is simulated using the same 
NRQCD action employed in recent studies of the $\Upsilon$ 
system \cite{upsilon}.
 For many of the ensembles the lattice spacing was 
determined from the $\Upsilon$ $2S-1S$ splitting.  For two ensembles where 
$\Upsilon$ results are not available, we used the 
heavy quark potential variable $r_1$ measured 
 by the MILC collaboration \cite{milc2,upsilon}. 
 The bare $s$ and $b$ 
quark masses have been fixed by the Kaon and $\Upsilon$ masses, respectively 
\cite{prl,upsilon} 
and based on  studies of light quark masses in \cite{lmass} we take 
as the physical chiral limit the point $m_s/m_q = 27.4$.

The basic quantity that needs to be calculated in decay constant 
determinations is the matrix element of the heavy-light axial vector 
current between the $B$ meson state and the hadronic vacuum.  Taking, as is 
customary, the temporal component of the axial current, in Euclidean 
space and in the $B$ rest frame one has
\be
\langle 0 | \, A_0 \, | B \rangle = M_B \, f_B .
\ee
In the last couple of years  we have made considerable 
progress in reducing statistical errors in numerical determinations
of this matrix element. 
We have developed better operators to create the $B$ meson state on 
the lattice and fit to 
a matrix of correlators with different smearings. Details of smearings 
and matrix fits are similar to those in the $\Upsilon$ spectroscopy studies 
of reference \cite{upsilon} and will not be repeated here.

  Table I summarizes results for the quantity 
$\Phi_q \equiv f_{B_q} \sqrt{M_{B_q}}$, where $B_q$ denotes a $''B''$ meson
with a light valence quark of mass $m_q$.
In the third column we show $a^{3/2} \, \Phi_q^{(0)}$,
 the result for $\Phi_q$ in 
lattice units when only the zeroth order lattice heavy-light current 
$J^{(0)} = \overline{\Psi}_q \, \gamma_5 \gamma_0 \, \Psi_Q$ is 
used.  The next column shows $a^{3/2} \, \Phi_q$, our results after 
one-loop matching and inclusion of $1/M$ currents. All corrections to 
the heavy-light current at 
${\cal O}(\Lambda_{QCD}/M)$, ${\cal O}(\alpha_s)$, 
${\cal O}(a \, \alpha_s)$, ${\cal O}(\alpha_s /(aM))$ and 
${\cal O}(\alpha_s \, \Lambda_{QCD}/M)$ have been included.
  The dimension 4 current corrections
 that enter into the matching at this order have been discussed 
in \cite{pert1}.
 The one-loop  perturbative matching coefficients specific to 
 the actions used in this study are given in \cite{pert2}.
 One sees that the 
difference between $\Phi_q^{(0)}$ and $\Phi_q$ is small, about 
$2 \sim 4$\% on the coarse lattices and $\sim 7$\% on the fine 
lattices.  The very small change on the coarse lattices may be 
partially accidental.  There is cancellation between the 
${\cal O}(\alpha_s)$ correction to the zeroth order current and 
the $1/M$ corrections.  The coefficient of the ${\cal O}(\alpha_s)$ 
term switches sign as one goes from a bare $b$ quark mass of 
$aM_0 = 2.8$ on the coarse lattices to $aM_0 = 1.95$ on the 
fine lattices, so that the cancellation does not occur on the  
latter. 
In the last column of Table I
we give results for $\Phi_q$ in ${\rm GeV}^{3/2}$. The first 
errors are statistical and the second come from lattice spacing uncertainties.
One sees that for most ensembles scale uncertainties dominate over 
statistical errors. The scales, $a^{-1}$, employed here are, in order of 
 the most chiral to the least chiral ensembles,
 1.623(32)GeV, 1.622(32)GeV, 1.596(30)GeV and 1.605(29)GeV, 
respectively on the four coarse lattices and 2.258(32)GeV and 2.312(31)GeV 
on the two fine lattices.

\begin{table}
\caption{ Simulation results for $\Phi_q \equiv f_{B_q} \sqrt{M_{B_q}}$.
Sea (valence) quark masses are denoted by $m_f$ ($m_q$) and 
$u_0 = [plaq]^{1/4}$ is the link variable used by the MILC collaboration 
in their normalisation of quark masses.
  See text for 
definitions of the last three columns.  The second error in the 
last column comes from uncertainties in the scale $a^{-3/2}$. 
}
\begin{center}
\begin{tabular}{|c|c|c|c|c|}
\hline
$u_0 \, a m_f$  & $u_0 \, a m_q$ & $\;\;a^{3/2} \Phi_q^{(0)}\;\;$ &
  $\;\;\;a^{3/2} \Phi_q \;\;\;$  &  $\;\Phi_q \, ({\rm GeV})^{3/2}\;$ \\
\hline
\hline
 Coarse  &&&&  \\
 0.005    & 0.005  & 0.2579(26)  & 0.2494(26) & 0.516(5)(15) \\
          & 0.040  & 0.3024(15)  & 0.2926(17) & 0.605(4)(18) \\
 0.007    & 0.007  & 0.2571(27)  & 0.2512(26) & 0.519(5)(15) \\
          & 0.040  & 0.2993(20)  & 0.2917(20) & 0.603(4)(18) \\
 0.010    & 0.005  & 0.2571(23)  & 0.2507(24) & 0.506(5)(14) \\
          & 0.010  & 0.2622(28)  & 0.2562(38) & 0.517(8)(15) \\
          & 0.020  & 0.2767(27)  & 0.2710(27) & 0.547(5)(15) \\
          & 0.040  & 0.3000(32)  & 0.2917(38) & 0.588(8)(17) \\
 0.020    & 0.020  & 0.2751(22)  & 0.2658(23) & 0.540(5)(15) \\
          & 0.040  & 0.2988(24)  & 0.2873(28) & 0.586(6)(16) \\
\hline
Fine  &&&&  \\
 0.0062   & 0.0062 & 0.1550(17)  & 0.1443(22) & 0.490(7)(10) \\
          & 0.031  & 0.1804(15)  & 0.1676(16) & 0.569(5)(12) \\
 0.0124   & 0.0124 & 0.1583(39)  & 0.1474(42) & 0.519(15)(10)\\
          & 0.031  & 0.1718(45)  & 0.1584(54) & 0.557(19)(11)  \\
\hline
\end{tabular}
\end{center}
\end{table}

\begin{figure}
\includegraphics[width=8.5cm,height=7.0cm]{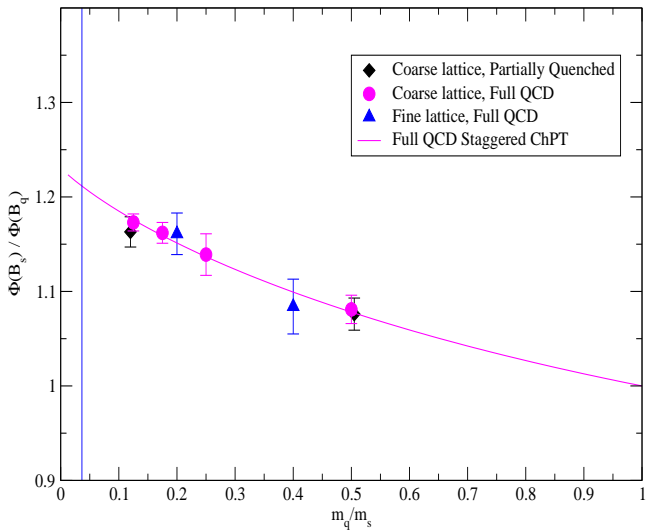}
\caption{The ratio $\xi_\Phi = \Phi_s/\Phi_q$ versus $m_q/m_s$. 
The full line through the data shows a fit to full QCD Staggered 
 $\chi PT$ (see text). Errors are statistical errors only.  The fine 
lattice points were not included in the fit.
 The vertical line at $m_q/m_s = 1/27.4$ 
denotes the physical chiral limit.
 }
\end{figure}

\begin{table}
\caption{ Simulation results for $\xi_\Phi \equiv \Phi_s/\Phi_q$ without
and with $1/M$ plus one-loop corrections.
}
\begin{center}
\begin{tabular}{|c|c|c|c|}
\hline
$u_0 \, a m_f$  & $u_0 \, a m_q$ & $\;\;\Phi_s^{(0)}/ \Phi_q^{(0)}\;\;$ &
  $\;\;\;\Phi_s / \Phi_q \;\;\;$   \\
\hline
\hline
 Coarse  &&&  \\
 0.005    & 0.005  & 1.173(7)  & 1.173(9) \\
 0.007    & 0.007  & 1.164(11) & 1.162(11) \\
 0.010    & 0.005  & 1.166(15) & 1.163(16) \\
          & 0.010  & 1.144(17) & 1.139(22) \\
          & 0.020  & 1.085(15) & 1.076(17)  \\
 0.020    & 0.020  & 1.086(13) & 1.081(15) \\
\hline
Fine  &&&  \\
 0.0062   & 0.0062 & 1.164(17) & 1.161(22)  \\
 0.0124   & 0.0124 & 1.092(19) & 1.084(29)  \\
\hline
\end{tabular}
\end{center}
\end{table}

Table II shows results for the ratio $\xi_\Phi \equiv \Phi_s/\Phi_q$.
  This quantity, unlike $\Phi_q$ itself, is not 
affected directly by errors in the lattice spacing.
Several other systematic errors inherent in $f_B$ determinations, that will
be discussed in more detail below, are also cancelled to a 
large extent in the ratio.
For instance, one sees that going from ratios of $\Phi^{(0)}$ to 
ratios of $\Phi$'s that include $1/M$ and one-loop matching corrections, 
produces almost no change at all.  
The data for $\xi_\Phi$ are plotted in Fig.1 as a function of $m_q/m_s$.  
The full curve comes from a fit to formulas of staggered chiral 
perturbation theory ($S\chi PT$) \cite{schpt1,schpt2,schpt3} and represents 
the prediction for full QCD.
The vertical line at small 
$m_q$ corresponds to the physical chiral limit $m_q/m_s = 1/27.4$.

$S\chi PT$ for heavy-light decay constants has been  developed 
by Aubin \& Bernard in reference \cite{schpt3}.  For $\Phi_q$ 
their formula reads,
\be
\label{phischpt}
\Phi_q = c_0 \; (1 \; + \; \Delta_q \;+ \; analytic).
\ee
The term encompassing the chiral logarithms, 
$\Delta_q \equiv \delta f_{Bq}/(16 \pi^2 f^2)$, is given in \cite{schpt3}
 and includes 
${\cal O}(a^2)$ effects coming 
from taste symmetry breaking, both in the mass splittings among 
light-light pseudoscalars and in lattice artifact hairpin diagrams.
For the ratio $\xi_{\Phi}$ we use the ansatz,
\be
\label{ansatz}
 \xi_\Phi = 1 + (\Delta_s -  \Delta_q) + \sum_k^{N_k} c_k
 \, (am_q - am_s)^k.
\ee
$N_k$ was increased until $\xi_{\Phi}^{(phys.)}$, the fit 
result for $\xi_{\Phi}$ at $m_q/m_s = 1/27.4$, and its error 
had stabilized (in practice $N_k = 2$ was sufficient).  Other ansaetze 
such as the direct ratio, 
$ \frac{1 + \Delta_s\, + \, c_1 \,  (2m_f+m_{sd}) 
\, + \, c_2 \, m_s}
{1 + \Delta_q\, + \, c_1 \, (2m_f+m_{sd}) 
\, + \, c_2 \, m_q}$
($m_{sd}$ is the sea $strange$ 
quark mass which, on the coarse lattices, is slightly larger than 
the true $strange$ quark mass $m_s$ we use for valence $strange$
 quarks) 
 or simple linear fits without any chiral logarithms 
were also tried as were fits with all the 
 ${\cal O}(a^2)$ taste breaking terms 
turned off.
 All these different chiral extrapolations 
 lead to values for $\xi_{\Phi}^{(phys.)}$ that 
differ at most by $3$\%. We fit simultaneously to the six coarse 
lattice points, 4 full QCD and 2 partially quenched (PQQCD) points, 
using full QCD and PQQCD $S \chi PT$ formulas respectively. Fig.1 shows 
just the full QCD curve.

The terms $\Delta_q$ involve the $BB^* \pi$ coupling $g_{B\pi}$ 
which is not known experimentally. We have carried out fits at several fixed 
values for $g_{B\pi}^2$ between $g_{B\pi}^2 = 0$ and $g_{B\pi}^2 = 0.75$. 
Good fits were obtained ($\chi^2/dof \approx 1$ or less) for $g_{B\pi}^2 
< 0.5$ with $\xi_\Phi^{(phys.)}$ differing again by less than $3$\% in the 
range $\xi_\Phi^{(phys.)}=1.21 \sim 1.24$.  We have also let $g_{B\pi}$ 
float as one of the fit parameters and find $g_{B\pi}^2 = 0.0(2)$ together with
$\xi_\Phi^{(phys.)} = 1.21(2)$. This fit result for $g_{B\pi}^2$ with 
the large uncertainty of $\Delta g_{B\pi}^2 = 0.2$ shows that 
 our data is not able to determine 
$g_{B\pi}^2$ with any accuracy, the same message we get from the 
fixed $g_{B\pi}$ fits, where a range of $g_{B\pi}^2$ 
between zero and $\sim 2 \times \Delta g_{B\pi}^2$  all give 
acceptable fits.  Fortunately, within this range $\xi_\Phi^{(phys.)}$ 
is not very sensitive to $g_{B\pi}^2$.  We take as our central value 
for $\xi_\Phi^{(phys.)}$ the result from the floating $g_{B\pi}$ 
fit, which we consider the least biased fit. This fit gives the curve shown on 
Fig.1.  We then take $\pm 0.03$ as the error due 
to statistics and chiral extrapolation uncertainties, and which also covers 
the spread we observe upon trying different ansaetze and 
different ways of handling $g_{B\pi}^2$.
 Remaining errors 
such as those due to discretization and relativistic corrections 
and higher order operator matchings not yet included, will affect 
$f_B$ and $f_{B_s}$ in similar ways and largely cancel in the ratio.
One expects their effects to come in at the level of the corresponding 
error in $\Phi_q$ times $a (m_s - m_q)$.  We have already seen that 
$1/M$ and one-loop matching corrections cancel almost completely 
in $\xi_\Phi$.  Furthermore the two full QCD fine lattice points in 
Fig.1 fall nicely on the full QCD $S \chi PT$ curve fixed by the 
coarse lattice points indicating that any residual 
discretization errors in $\xi_\Phi$ are smaller than the current statistical
errors.  Taking all these arguments into account,
we estimate a $\sim 1$\% further uncertainty in $\xi_\Phi$ from 
these other sources. Our final result for $f_{B_s}/f_B =
 \xi_\Phi \, \sqrt{\frac{M_B}{M_{B_s}}}$ is then
\be
\label{fbrat}
f_{B_s} / f_B = 1.20(3)(1) .
\ee
We emphasize that the reason the chiral extrapolation errors 
are small here is because the light quark action 
employed in this study allowed us to go down as low as $m_s/8$ and 
only a modest extrapolation to the physical chiral limit was required.
This differs from the case with 
Wilson type light quarks, where simulations have typically been restricted 
to $m_q/m_s > 0.5$, i.e. to the region to the
 right of the heaviest data point in Fig.1.

\begin{figure}
\includegraphics[width=8.5cm,height=7.0cm]{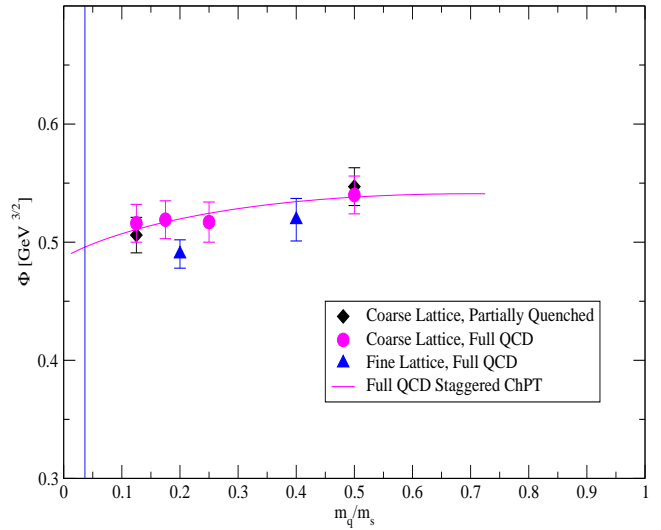}
\caption{$\Phi_q$ versus $m_q/m_s$.  Errors include both statistical and 
scale uncertainty errors. The fine lattice points were not 
included in the fit.
 }
\end{figure}

Fig.2 shows the data points for 
$\Phi_q$ itself for $m_q/m_s \leq 0.5$ together with a full QCD $S \chi PT$ 
fit curve.
For chiral extrapolation of $\Phi_q$  we use directly 
eq.(\ref{phischpt}) with analytic terms $c_1 (2 m_f+m_{sd}) + c_2 m_q$. 
We again carry out simultaneous fits to the coarse lattice 
full QCD and PQQCD points.
Fits with the coupling $g_{B\pi}^2$ held fixed between 
0.0 and 0.6 all lead to good fits 
with $\Phi^{(phys.)}$ varying by $4$\%.  Allowing this coupling 
to float gives $g_{B\pi}^2 = 0.1(5)$,
 which is consistent with the fixed $g_{B\pi}$ fit results, 
and $\Phi^{(phys.)} = 0.496(20)$
GeV$^{3/2}$ with again a $4$\% error. We take the $4$\% to be our best 
estimate for the combined error from
 statistics, chiral extrapolation and determination of $a^{-1}$.
 The full QCD $S \chi PT$ curve in Fig.2 comes 
from the floating $g_{B\pi}^2$ fit.  
 We turn next to estimates of the other systematic errors in $\Phi^{(phys.)}$.

A major source of systematic error in $\Phi^{(phys.)}$ is higher order
 matching of the 
heavy-light current.  Although the one-loop contributions turned 
out to be small (as described above), in fact much 
smaller than a naive estimate of  ${\cal O}(\alpha_s) \sim 30$\%, 
we have no argument guaranteeing 
this to be true at higher orders.  Hence we allow for an ${\cal O}(\alpha_s^2) 
\approx 9$\% systematic matching error. This will be the dominant 
systematic error in our decay constant determination. Another source of 
systematic error comes from discretization effects. 
The fine lattice points in Fig.2 lie about $3 \sim 5$\% lower than those from
the coarse lattices.   Since the statistical plus scale uncertainty errors 
on all our points range between $2 \sim 3$\%, it is not obvious how much of
this difference comes from discretization effects.
The size of fluctuations between independent coarse ensembles is 
comparable to this difference.  It should also be noted that 
the difference between the coarse and fine lattice data would disappear 
if it were not for the one-loop matching corrections (recall the 
$2 \sim 4$\% corrections on the coarse lattices versus the 
$\sim 7$\% corrections on the fine lattices giving a $3 \sim 5$\% 
difference in the radiative corrections on the two lattices). In other 
words it is difficult to
disentangle discretization errors from radiative corrections. 
One could quote a combined discretization and higher order matching 
error again at the $\sim 9$\% level.  We opt instead to keep 
the $9$\% as the pure (and dominating) ${\cal O}(\alpha_s^2)$ error and 
use a conventional naive estimate of ${\cal O}(a^2 \alpha_s) \approx 
2$\%  for discretization errors.  As the last nontrivial systematic error
we estimate uncertainties from relativistic corrections and tuning of the 
$b$ quark mass \cite{upsilon}
 to be at the $\sim 3$\% level.  Putting all this together 
we obtain $\Phi^{(phys.)} = 0.496(20)(45)(10)(15) \, {\rm GeV}^{3/2}$. 
 This leads to our result for the $B$ meson decay constant of
\be
\label{fb}
f_B = 0.216(9)(19)(4)(6) \; {\rm GeV}.
\ee
The errors, from left to right, come from statistics plus scale plus 
chiral extrapolations, higher order matching, discretization, and 
relativistic corrections plus $m_b$ tuning
 respectively. Combining this result with 
our result for $f_{B_s}/f_B$, eq.(\ref{fbrat}), 
 one finds $f_{B_s} = 0.259(32)\,$GeV. This 
is very consistent with the direct calculation of 
$f_{B_s}$   published earlier in \cite{fbsprl} where we quote a 
 value of $0.260(29) \, $GeV.

To summarize, we have completed a determination of the $B$ meson 
 decay constant in full (unquenched) QCD.  Our main results are given in eqns.
(\ref{fbrat}) and (\ref{fb}). The use of a highly improved light quark 
action has led to good control over the chiral extrapolation to 
physical $up$ and $down$ quarks. Better smearings have significantly 
reduced statistical errors. 
For the ratio $f_{B_s}/f_B$ these 
improvements  translate into an accurate final result with errors at the
 $\sim 3$\% level.
For $f_B$ itself other systematic errors 
not yet addressed in the present study dominate and the 
current total error is at the $\sim10$\% level. The main remaining 
source of uncertainty comes from higher order operator matching.  More studies 
should also be carried out on the fine lattices and on even finer
lattices currently being created by the MILC collaboration,  to reduce 
discretization uncertainties. Errors in the scale $a^{-1}$ 
need to come down for all the ensembles.
 Improvements on all these fronts are underway.  Calculations of the 
bag parameter $B_B$ have also been initiated. 

This work was supported by the DOE and NSF (USA) and by PPARC (UK).
 A.G., J.S. and M.W. thank the KITP 
U.C. Santa Barbara for support during the workshop,
 ``Modern Challenges in Lattice Field Theory'' when part of the present 
research was carried out.
Simulations were done at NERSC and on the 
Fermilab LQCD cluster.  We thank Steve Gottlieb and the MILC collaboration 
for making their dynamical gauge configurations available.
 We are also 
grateful to Jim Simone and the Fermilab collaboration for use of 
their light propagators on the fine lattices and to Claude Bernard 
for sending us his notes on $S\chi PT$ for heavy-light decay constants.



\end{document}